\documentclass[journal,transmag]{IEEEtran}
\usepackage{cite}
\usepackage{amsmath,amssymb,amsfonts}
\usepackage{algorithmic}
\usepackage{graphicx}
\usepackage{textcomp}
\usepackage{url}

\def\BibTeX{{\rm B\kern-.05em{\sc i\kern-.025em b}\kern-.08em
    T\kern-.1667em\lower.7ex\hbox{E}\kern-.125emX}}

\begin{document}
\bibliographystyle{IEEEtran}

\title{Optimized Feedforward Neural Network Training for Efficient Brillouin Frequency Shift Retrieval in Fiber}
\author{\IEEEauthorblockN{Yongxin Liang\IEEEauthorrefmark{1},
		Jialin Jiang\IEEEauthorrefmark{1},
		Yongxiang Chen\IEEEauthorrefmark{1}, 
		Richeng Zhu\IEEEauthorrefmark{1}, 
		Chongyu Lu\IEEEauthorrefmark{1}, 
		and	Zinan Wang\IEEEauthorrefmark{1, 2}, \IEEEmembership{Senior Fellow, IEEE}}

	\thanks{\IEEEauthorrefmark{1}Key Lab of Optical Fiber Sensing and Communications, University of Electronic Science and Technology of China, Chengdu, Sichuan, China 611731. 
		
		\IEEEauthorrefmark{2}Center for Information Geoscience, University of Electronic Science and Technology of China, Chengdu, Sichuan, China 611731
		
		This work is supported by Natural Science Foundation of China (41527805, 61731006), Sichuan Youth Science and Technology Foundation (2016JQ0034), Guofang Keji Chuangxin Tequ, and the 111 project (B14039).
		
		 Corresponding author: Zinan Wang (e-mail: znwang@uestc.edu.cn).}}

\IEEEtitleabstractindextext{%
	\begin{abstract}
		Artificial neural networks (ANNs) can be used to replace traditional methods in various fields, making signal processing more efficient and meeting the real-time processing requirements of the Internet of Things (IoT). As a special type of ANN, recently the feedforward neural network (FNN) has been used to replace the time-consuming Lorentzian curve fitting (LCF) method  in Brillouin optical time-domain analysis  (BOTDA) to retrieve the Brillouin frequency shift (BFS), which could be used as the indicator in temperature/strain sensing, etc. However, FNN needs to be re-trained if the generalization ability is not satisfactory, or the frequency scanning step is changing in the experiment. This is a cumbersome and inefficient process. In this paper, FNN only needs to be trained once with the proposed method. 150.62 km BOTDA is built to verify the performance of the trained FNN. Simulation and experimental results show that the proposed method is promising in BOTDA because of its high computational efficiency and wide adaptability.
	\end{abstract}
	
	\begin{IEEEkeywords}
	Feedforward neural networks, Brillouin optical time-domain analysis, Lorentzian curve fitting,  Sensor fusion,  Optical fiber sensors.
\end{IEEEkeywords}}

\maketitle

\section{Introduction}
\label{sec:introduction}
The Internet of Things (IoT) is constantly progressing as multiple technologies are evolving, especially the sensing technologies\cite{1,2,3,4,5}. 
As an important branch of sensors, optical fiber sensors,  have been studied by many researchers due to their unique advantages\cite{6}. 
Particularly, distributed optical fiber sensing (DOFS) systems are of great interests since they can turn fiber cables into massive sensor arrays\cite{7,8,9,10}. 
Brillouin optical time-domain analysis (BOTDA) system is an important type of DOFS, which could achieve high precision, long distance and fast scan-rate sensing \cite{11,12,13,14}. 

The stimulated Brillouin scattering (SBS) effect is the basis of BOTDA\cite{15,16}. 
In BOTDA, usually pump pulse and continuous-wave (CW) probe light are counter-propagating inside the fiber to sample the Brillouin gain spectrum (BGS), then the Brillouin frequency shift (BFS) is retrieved for the purpose of temperature/strain sensing. 
Before the traditional Lorentzian curve fitting (LCF) method is used to find the BFS\cite{17}, the non-local means (NLM), wavelet denoising (WD) or block-matching and 3D filtering  (BM3D)   can be used to reduce the noise of the BGS in general\cite{18}. 
All of those processes are time-consuming operations, especially for longer sensing distance and finer spatial resolution. 
Recently, denoise convolutional neural network (DnCNN) is used for BOTDA filtering \cite{19}, which can achieve real-time filtering as long as the DnCNN is trained properly. 
However, the training process of DnCNN is complicated.

On the other hand, artificial neural networks(ANNs) have been applied in BOTDA\cite{22,23,24}. 
As a special type of ANN, the feedforward neural network (FNN) can be used to replace the traditional LCF method and improve the processing speed.  
Ideal BGSs are used for FNN training\cite{22}. 
During FNN testing process, the noise in the measured BGS needs to be reduced as much as possible, which means that the filtering operation cannot be omitted. 
However, using traditional filtering methods will not meet the requirements of real-time analysis. 
Although using DnCNN can achieve fast filtering, the training process is complicated. Without filtering, it means that FNN trained by ideal  BGSs may not have generalization ability for the noisy BGS in the experiment. 
Regularization method is needed in order to  enhance the generalization ability of FNN\cite{25,26,27}. 
This is a time-consuming task.

Also, the frequency scanning step is one of the key parameters in BOTDA operation. 
It usually varies between 1 MHz and 10 MHz depending on the required accuracy and speed of the measurement. 
Previous work required training different FNNs to meet the needs of different frequency scanning steps, which is a tedious task.

In this paper,  optimized FNN training method is proposed for BOTDA.
 As a result, BFS retrieval from the BGSs can be efficiently carried out with a wide range of BGS linewidths, signal-to-noise ratios (SNRs) and frequency scanning steps. 
 The results show that the accuracy of FNN is similar to  LCF, while  FNN is much faster than LCF. 
 The 23.95 km and 150.62 km BOTDA at the frequency step of 1 MHz and 4 MHz are established respectively to experimentally verify the performance of FNN.

This paper is organized as follows.
 Section II describes the principles of LCF and FNN to get BFS from BGSs. 
 Section III describes  FNN training process and the results. 
 Section IV explains how to enhance the adaptability of the trained FNN. 
 Section V presents the experiment and analyses the results.  
 Section VI gives the conclusion. 

\section{The principles of LCF and FNN}
The principles for BFS retrieval from the BGSs using LCF and FNN are explained respectively in this section.

\subsection{The principle of LCF}
In BOTDA, the obtained local BGS conforms to the shape of Lorenz curve. 
Equation \eqref{eq1} is the core of BGSs simulation.
\begin{equation}{g}(v,z)=\frac{{{G}_{B}}(z)}{1+{{[2(v-{{v}_{B}})/\Delta {{v}_{B}}]}^{2}}}.\label{eq1}\end{equation}
where $\Delta v_B$ is the linewidth. 
$z$ is the position of the fiber. 
${G}_{B} (z)$  is the gain coefficient.  
$v$   is the scanning frequency.  
${g}(v,z)$ is the measured BGS. 
${v}_{B}$ is BFS which reflects information about temperature or strain. 
The fiber is placed in a stable environment so that the change of  BFS is only caused by temperature theoretically:

\begin{equation}{{v}_{B\text{2}}}-{{v}_{B\text{1}}}={{C}_{T}}({{T}_{\text{2}}}-{{T}_{\text{1}}}).\label{eq2}\end{equation}
where ${{C}_{T}}$ is the temperature coefficient. 
Temperature difference can be obtained by measuring the difference of BFS. 
In order to obtain BFS, the least squares estimation (LSE) method is used in LCF.

\begin{equation}{{\text{R}}^{2}}\text{=}{{\sum\nolimits_{i=1}^{N}{[g({{v}_{i}},z)-{{y}_{i}}]}}^{2}}.\label{eq3}\end{equation}

The purpose of (3) is to minimize the ${{\text{R}}^{2}}$ by modifying parameters during iterations. 
At the end of the iteration, y is the result of LCF. 

\subsection{The principle of FNN}
The hidden layer of FNN is set according to the size of training data and the number of input and output nodes. 
A typical FNN is showed in Fig 1.
\begin{figure}[htb!]
	\centering
	\includegraphics[scale=1]{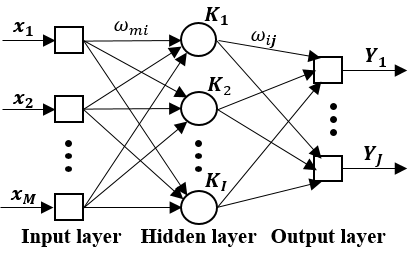}
	\caption{Typical FNN with three layers.}
	\label{fig1}
\end{figure}

The number of the input layer, the hidden layer, and the output layer are ${M}$, ${I}$ and ${J}$, respectively. 
The input and output of the FNN are ${{X}_{m}}$ and ${{Y}_{j}}$, respectively. 
The input and output of the hidden layer ${{K}_{i}}$ are ${{u}_{i}}$ and ${{v}_{i}}$, respectively. 
The weight from ${{X}_{m}}$ to ${{K}_{i}}$ and ${{K}_{i}}$ to ${{Y}_{j}}$ are ${{\omega }_{mi}}$ and ${{\omega }_{ij}}$, respectively. 
The FNN accepts a vector of length ${M}$ as the input data, and finally generates a vector of length ${J}$ as the output data. 
The input of the hidden layer ${{K}_{i}}$ in the  ${n}^{th}$ iteration is:
\begin{equation}{{u}_{i}}(n)=\sum\nolimits_{m=1}^{M}{{{\omega }_{mi}}(n)}{{x}_{m}}(n).\label{eq4}\end{equation}

The output of the hidden layer ${{K}_{i}}$ is:
\begin{equation}{{v}_{i}}(n)=f({{u}_{i}}(n)).\label{eq5}\end{equation}

where $f($·$)$ is the sigmoid function. The output of the network is:
\begin{equation}	{{y}_{j}}(n)=\sum\nolimits_{i=1}^{I}{{{\omega }_{ij}}(n)}{{v}_{i}}(n).\label{eq6}\end{equation}
\begin{equation}Y(n)=[{{y}_{1}}(n),{{y}_{2}}(n),...,{{y}_{J}}(n)].\label{eq7}\end{equation}

The expected output of the network is:
\begin{equation}H(n)=[{{h}_{1}}(n),{{h}_{2}}(n),...,{{h}_{J}}(n)].\label{eq8}\end{equation}

The error signal in the ${n}^{th}$ iteration is defined as:
\begin{equation}{{e}_{j}}(n)={{h}_{j}}(n)-{{y}_{j}}(n).\label{eq9}\end{equation}

And the cost function can be defined as:
\begin{equation}e(n)=\frac{1}{2J}\sum\nolimits_{j=1}^{J}{e_{j}^{2}(n)}.\label{eq10}\end{equation}

In order to minimize the error, the basic method is to use the steepest descent algorithm for backpropagation (BP) calculations, which modifies the weight according to \eqref{eq11} and \eqref{eq12}. Besides, there are other algorithms, such as conjugate gradient algorithm, Levenberg-Marquardt(LM) algorithm and so on.
\begin{equation}\Delta {{\omega }_{ij}}(n)=-\eta \frac{\partial e(n)}{\partial {{\omega }_{ij}}}.\label{eq11}\end{equation}
\begin{equation}{{\omega }_{ij}}(n+1)=\Delta {{\omega }_{ij}}(n)+{{\omega }_{ij}}(n).\label{eq12}\end{equation} 

 The input vector is the local BGS, and the output vector is BFS of the local BGS. 
 It should be noted that both the input local BGS and the output BFS need to be normalized. 
 The reason for the input local BGS needed to be normalized is that as the sensing distance increases, the gain of the local BGS decreases, and the normalization operation eliminates the effects of different gains. 
 Besides, it is convenient to calculate the actual BFS in different parameters among experiments when the output BFS is normalized. 
 The training process of FNN is showed in Fig. 2.

\begin{figure}[htb!]
	\centering
	\includegraphics[scale=0.687]{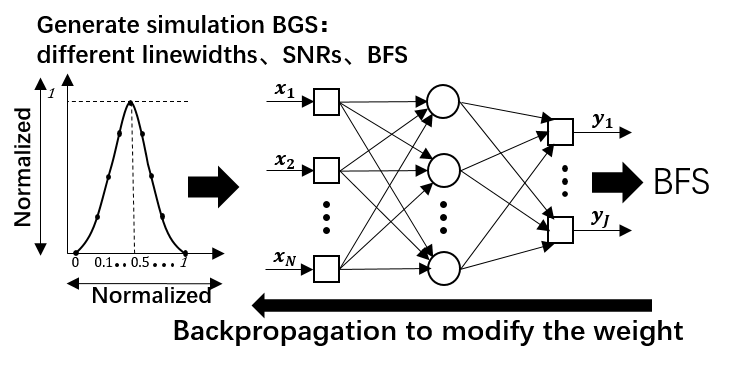}
	\caption{FNN training process.}
	\label{fig2}
\end{figure}

This section provides the principles for BFS retrieval by these two methods. 
As a traditional method, LCF has a guaranteed accuracy but a time-consuming process, while FNN has the ability to calculate quickly but requires verification of accuracy.

\section{OPTIMIZED FNN TRAINING: SIMULATION RESULTS}
Training a FNN with good generalization ability is not an easy task. 
Using the optimized training process of adding noise to ideal BGSs method can improve the generalization ability of  FNN\cite{28}.  It will be described in this section.

Generally, in BOTDA, the measured BGS needs to be scanned in a wide frequency range to find the location of BFS. 
Here, 156 MHz frequency scanning range at the frequency scanning step of 1 MHz is selected so that the input data of the neural network is a vector of length 157. 
The layout of hidden layers is set to 40-15 after several attempts. 
Therefore, the layout of FNN is 157-40-15-1. 
The size of simulated BGSs is 157×385560, which satisfies the linewidths range from 10 MHz to 60 MHz at a step of 1 MHz, and the BFS varies from 10\% to 90\% in the 156 MHz frequency range at a step of 1 MHz. 
In addition, it also contains BGSs under different power noises.

The addition of random noise to training data can enhance the generalization ability of FNN. 
Therefore, the simulated BGSs with several SNRs are generated to train FNN. 
The range of SNRs is 16 dB to 36 dB at a step of 10 dB. 
20 sets of random noise are added to the ideal BGS in each SNR. 
Three samples are selected from all simulated BGSs for observation and showed in Fig. 3.

\begin{figure}[htb!]
	\centering
	\includegraphics[scale=0.555]{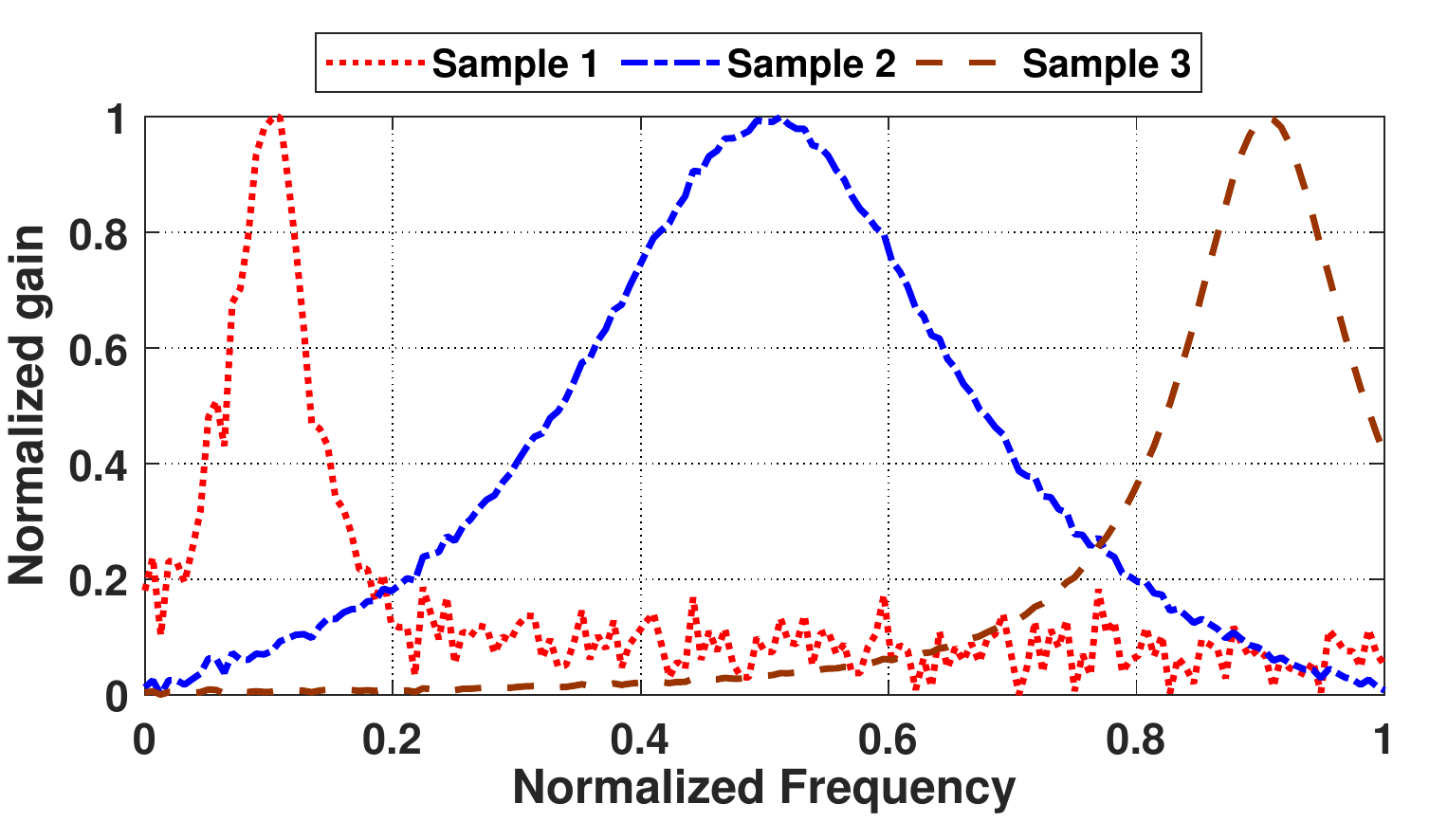}
	\caption{The simulated BGSs with different linewidths, SNRs and BFS. (Sample 1: SNR=16 dB, linewidth=10 MHz, ${{v}_{B}}$=10\%. Sample 2: SNR=30 dB, linewidth=60 MHz, ${{v}_{B}}$=50\%. Sample 3: SNR=45 dB, linewidth=25 MHz, ${{v}_{B}}$=90\%).}
	\label{fig3}
\end{figure}

The simulated 16 dB BGSs with different linewidths are used as the test dataset in FNN training process. 
Using the LM training algorithm, the training process takes about 9 hours in 30 iterations. 
As shown in Fig. 4, the error of the training and test dataset has the same order of magnitude at the  $30^{th}$ iteration, which means that the trained FNN has generalization ability for the 16 dB BGSs.

\begin{figure}[htb!]
	\centering
	\includegraphics[scale=0.595]{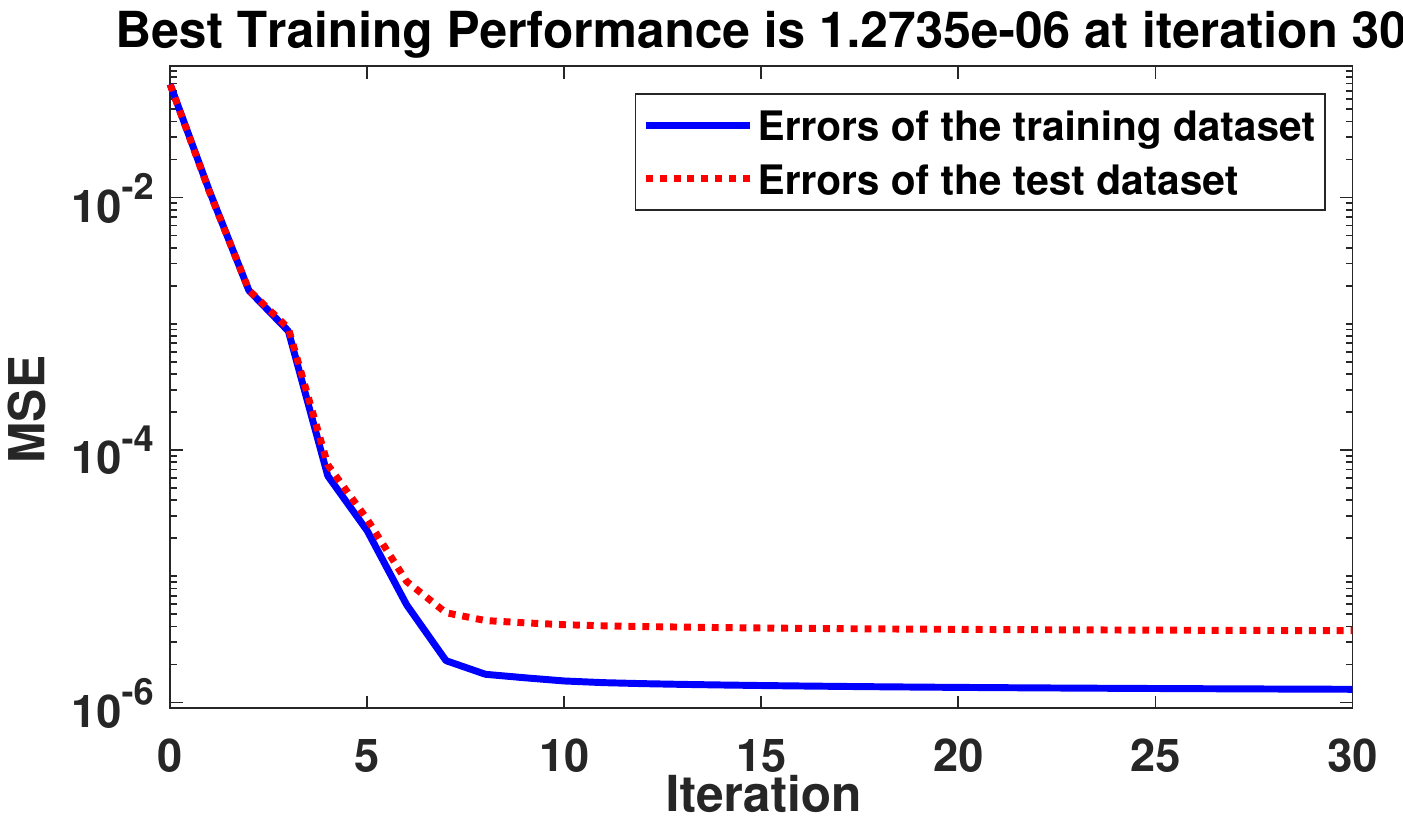}
	\caption{The mean squared error (MSE) of each iteration of the FNN trained by the noisy BGSs.}
	\label{fig4}
\end{figure}

For comparison, the simulated BGSs without noise are used as the training dataset. 
Other simulation and training parameters are consistent, and the error results are showed in Fig. 5. 
The error of the training and test dataset has different orders of magnitude at the  $30^{th}$ iteration, which means that the trained FNN cannot retrieve the BFS from the 16 dB BGSs accurately. 
The error of the training dataset is driven to a small value while the error of the test dataset still in a high level, which means the overfitting occurs.

\begin{figure}[htb!]
	\centering
	\includegraphics[scale=0.6]{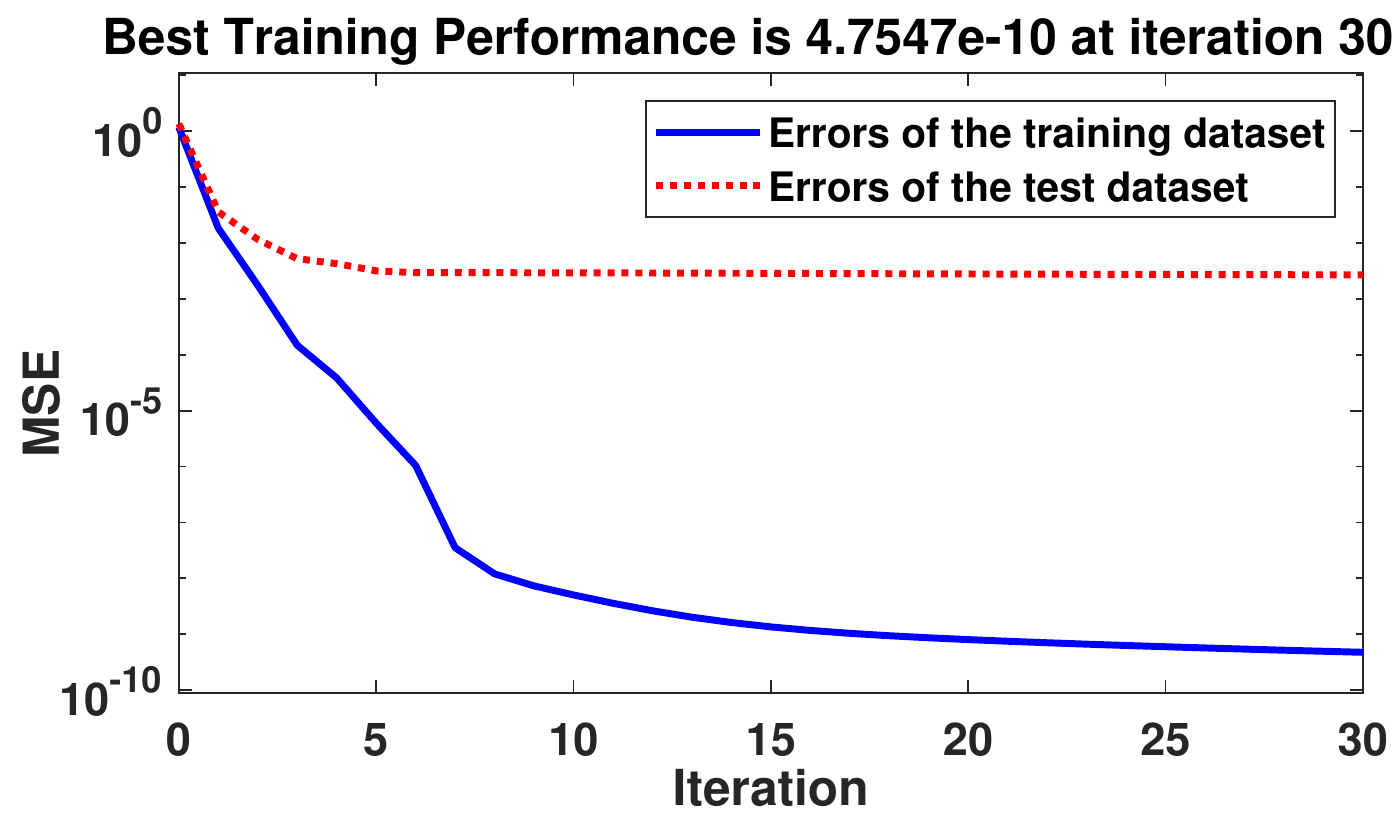}
	\caption{The MSE of each iteration of the FNN trained by the ideal BGSs.}
	\label{fig5}
\end{figure}

Through this optimized training process of noise addition, the trained FNN has good generalization ability and greatly improves the training efficiency without additional regularization methods. 
The trained FNN can be used to calculate BFS directly. 

Next, the simulated BGSs with different SNRs from 16 dB to 46 dB at a step of 1 dB are used for test. 
The root-mean-square error (RMSE) of BFS is calculated by FNN and LCF, respectively. 
The result is showed in Fig. 6.

\begin{figure}[htb!]
	\centering
	\includegraphics[scale=0.581]{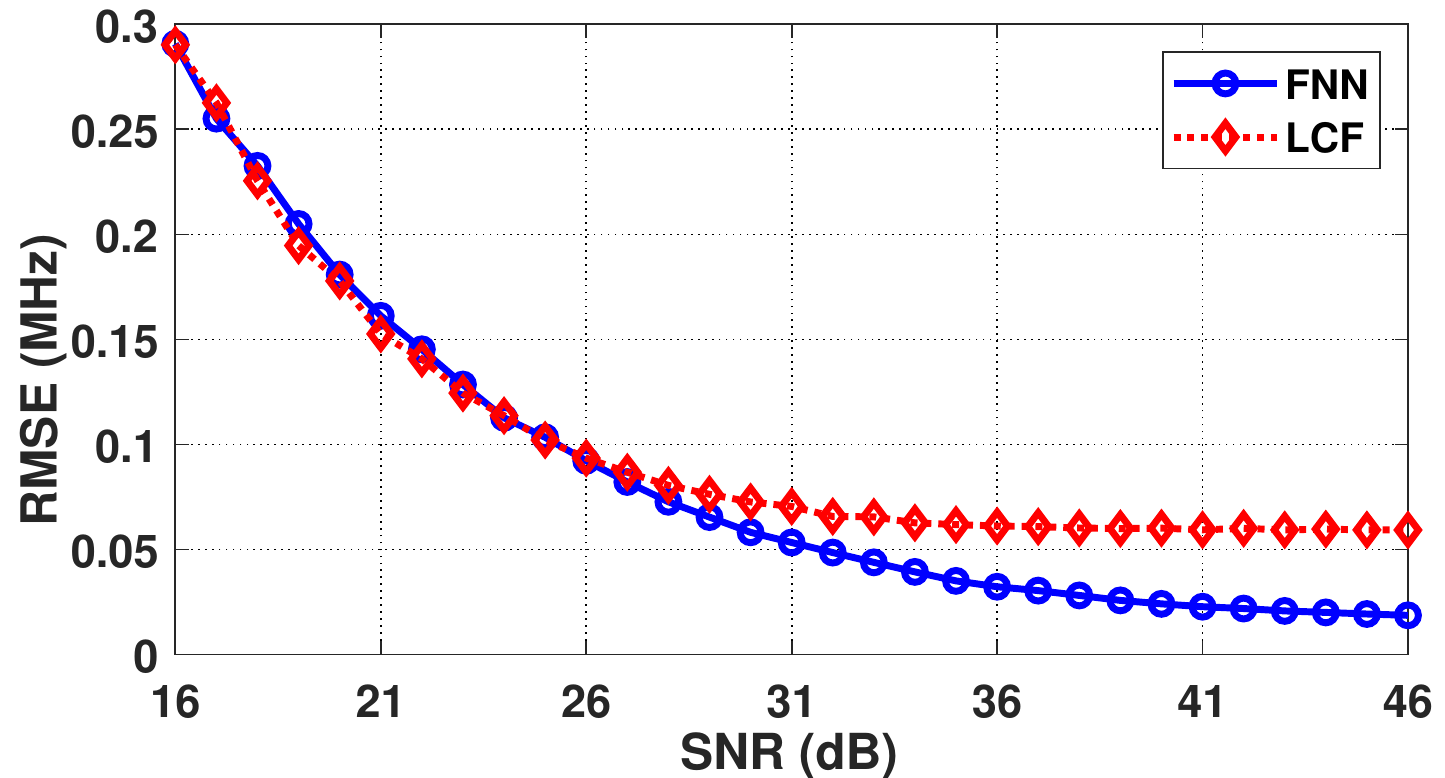}
	\caption{The RMSE of FNN and LCF for the simulated BGSs with different SNRs.}
	\label{fig6}
\end{figure}

Fig. 6 shows that the RMSE of BFS calculated by  FNN and LCF from  the BGS with lower SNR is similar.  
With the increase of SNR, the RMSE calculated by FNN can be reduced to below 0.05 MHz, while the error of LCF is always above 0.05 MHz. 
Besides, the simulated 16 dB BGSs with different linewidths at a frequency scanning step of 1 MHz are used for test, as shown in Fig. 7.

\begin{figure}[htb!]
	\centering
	\includegraphics[scale=0.582]{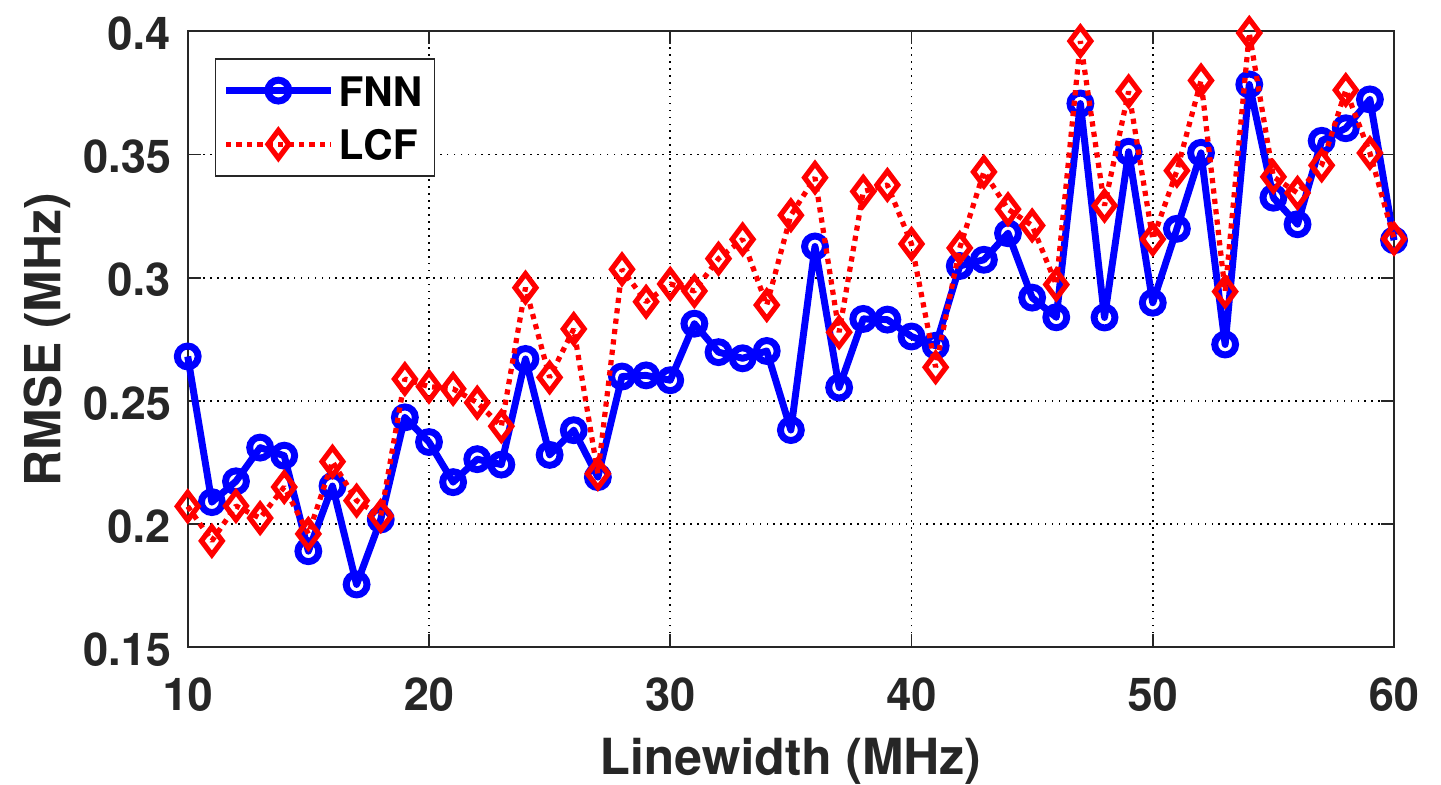}
	\caption{The RMSE of FNN and LCF for the simulated 16 dB BGSs with different linewidths.}
	\label{fig7}
\end{figure}

Through the optimized training method, a FNN that can calculate BFS accurately from the BGSs with different linewidths and SNRs is trained.

\section{THE FNN ADAPTABILITY: SIMULATION RESULTS}
The FNN trained in the previous section will not be available if the frequency scanning step is not 1 MHz in the experiment, which means that the BGSs at the corresponding frequency scanning step are needed to be simulated again and used as the training dataset to train another FNN. 
In long distance BOTDA, it is more desirable to use a higher frequency scanning step, such as 4 MHz. 
This will increase the measured speed of the BGS at the expense of the accuracy of the temperature measurement. 
Correspondingly, a lower frequency scanning step, such as 1 MHz, is selected under high precision requirements. 
Training different FNNs using the BGSs at different frequency scanning steps can be used to solve this issues.

However, the FNN training is a cumbersome process that involves generating simulated data, adjusting hidden layers and other parameters until the FNN has generalization ability and  satisfactory error results. 
Therefore, it will be convenient if only one FNN needs to be trained, which can retrieve  BFS from  BGS at different frequency scanning steps. 
In order to solve this issue, the linear interpolation method can be used to reshape the other frequency scanning step of the BGS to 1 MHz. 

Consistency is required to obtain accurate results. 
The frequency scanning range should not be lower than 156 MHz because the previous FNN is trained using simulated BGSs with the frequency scanning range of 156 MHz. 
In order to ensure the frequency scanning range is an integer multiple of the frequency scanning step, the minimum frequency scanning ranges at different frequency scanning steps are different. 
They are at least 156 MHz, 156 MHz, 156 MHz, 156 MHz, 160 MHz, 156 MHz, 161 MHz, 160 MHz, 162 MHz, 160 MHz, corresponding to different frequency scanning steps from 1 MHz to 10 MHz at a step of 1 MHz. 
The linear interpolation is used to reshape the BGS to the frequency scanning step of 1 MHz. 
At last, the interpolated BGS in appropriate frequency scanning range of 156 MHz is selected and input into the FNN to retrieve the BFS.

The BFS is retrieved by FNN and LCF from the 16 dB BGSs with different linewidths at the different frequency scanning steps, and the result of the RMSE is showed in Fig. 8. 
The error of  FNN and LCF  is basically same. 
Both of them are small compared with the measurement uncertainty in the experiment.

\begin{figure}[htb!]
	\centering
	\includegraphics[scale=0.575]{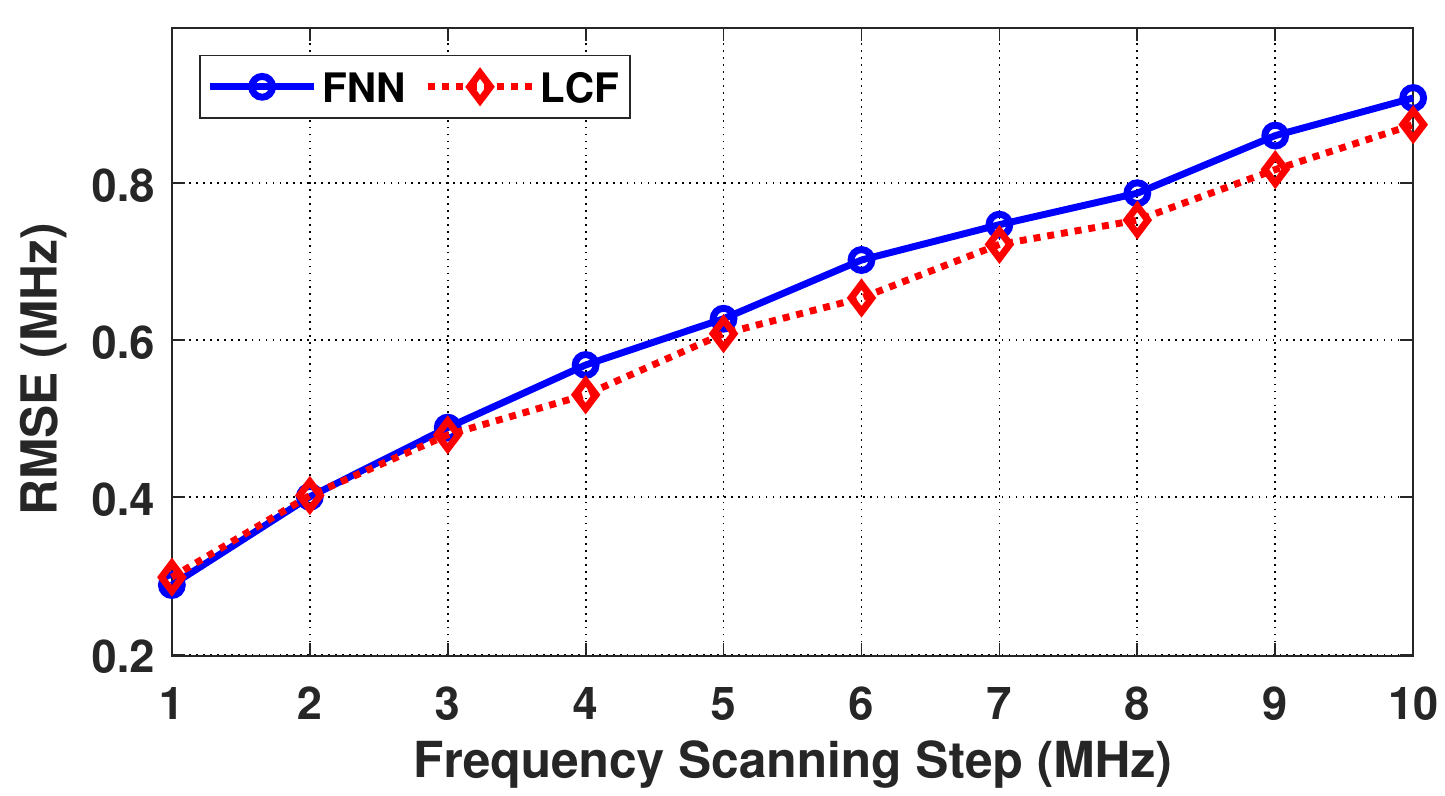}
	\caption{The RMSE of FNN and LCF for the 16 dB BGSs with different linewidths at different frequency scanning steps.}
	\label{fig8}
\end{figure}

In order to compare efficiency, the BFS is retrieved by  FNN and LCF from 10,000 sets of the local BGS at each frequency scanning step.
 The consumption time ratios of LCF (single thread and 16 threads) to FNN (single thread) is showed in Fig. 9.

\begin{figure}[htb!]
	\centering
	\includegraphics[scale=0.575]{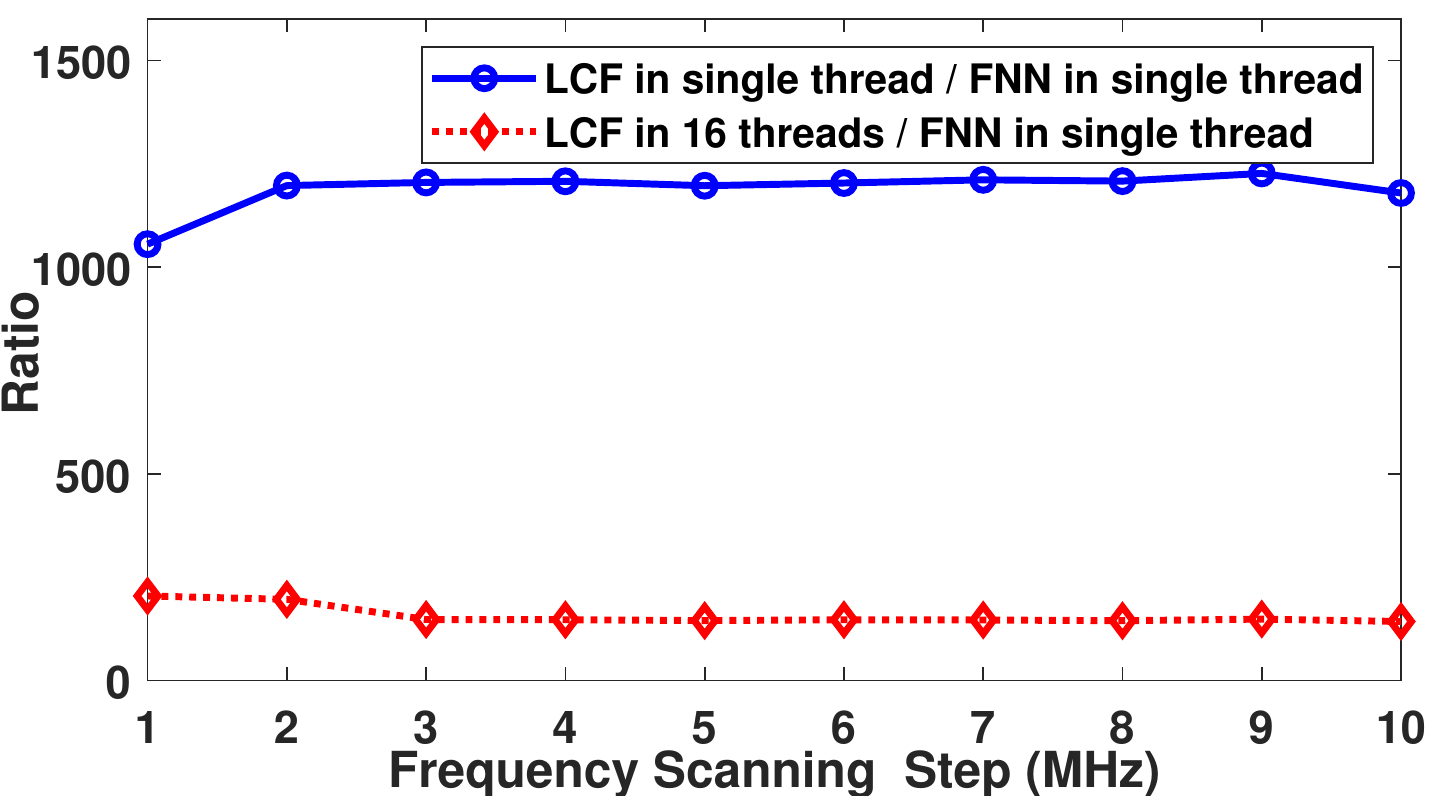}
	\caption{The ratios of LCF to FNN in different frequency scanning steps.}
	\label{fig9}
\end{figure}

The interpolation method enables BGSs at different frequency scanning steps to be input into a trained FNN and then obtain the accurate BFS calculation. 
The high efficiency and wide adaptability of FNN will promote the development of BOTDA.

\section{EXPERIMENTAL DEMONSTRATION}
In order to analyze the ability of the FNN for calculating the BFS from the BGS in the experiment, the 23.95 km BOTDA with 200 MHz frequency range at a frequency scanning step of 1 MHz and 150.62 km BOTDA with 156 MHz frequency range at a frequency scanning step of 4 MHz are established, respectively.

\subsection{Experimental setup of the 23.95 km BOTDA}
The experimental setup of the 23.95 km BOTDA is showed in Fig. 10. 
 This is a typical BOTDA experimental system. 
 Pump pulse and CW probe light are counter-propagating inside the fiber under test (FUT) to sample the BGS.
  The BGS is measured by direct detection in the frequency scanning range of 200 MHz, with the frequency scanning step of 1 MHz.
\begin{figure}[htb!]
	\centering
	\includegraphics[scale=0.75]{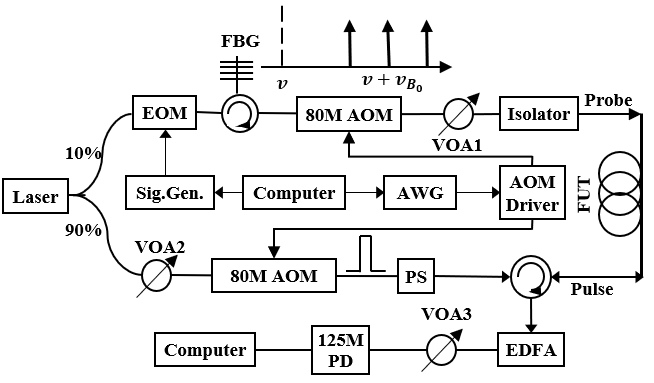}
	\caption{Experimental setup of 23.95 km BOTDA. EOM: electric-optical modulator; FBG: fiber Bragg grating; AOM: acoustic optical modulator; PS: polarization scrambler; EDFA: erbium-doped fiber  amplifier; AWG: arbitrary waveform generator; VOA: adjustable attenuator.}
	\label{fig10}
\end{figure}

The heating location is at 23.7 km. The SNR is 23.5 dB, approximately. 
As can be seen from Fig. 11, the position of the BFS is changed due to heating.

\begin{figure}[htb!]
	\centering
	\includegraphics[scale=0.55]{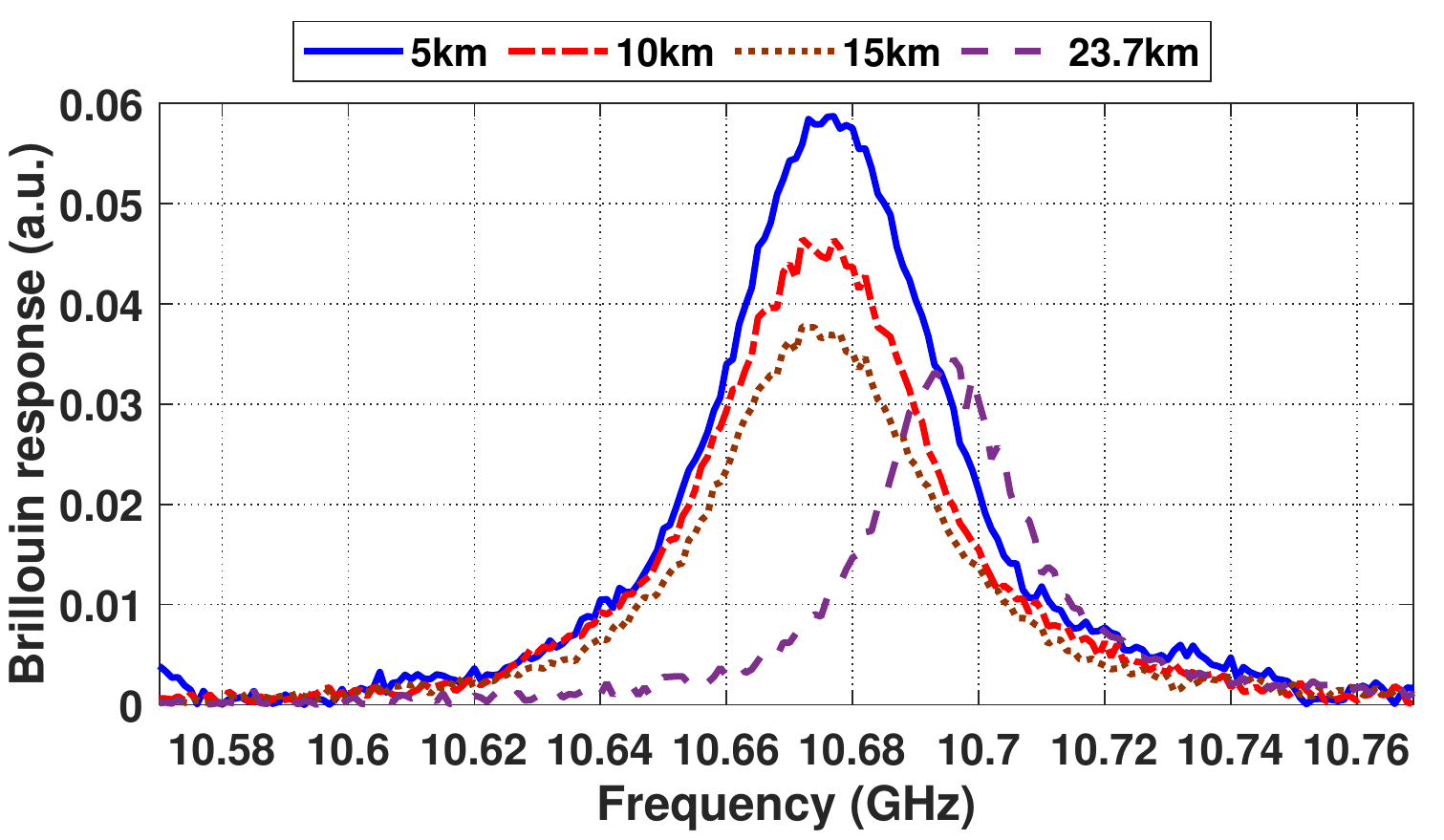}
	\caption{local BGSs at different locations of 23.95 km BOTDA.}
	\label{fig11}
\end{figure}

The input of the FNN is the local BGS in the frequency scanning range of 156 MHz at a frequency scanning step of 1 MHz. 
However, the frequency scanning range of the measured BGS is 200 MHz at a frequency scanning step of 1 MHz. 
 Therefore, the BGS with appropriate 156 MHz frequency range is selected according to the position of the power peak and input to the neural network to retrieve the BFS. 
 Also, the BFS is calculated by LCF for comparison. Subtracting BFS before and after heating to obtain the frequency difference, which reflects the change of temperature, as shown in Fig. 12. 
 The temperature coefficient is 1.3 MHz/$^\circ$C and the applied temperature difference is 15.7$^\circ$C. 
 The temperature difference measured by LCF and FNN are both 15.6$^\circ$C. 
 The measurement uncertainty calculated by LCF and FNN are $\pm$0.22$^\circ$C and $\pm$0.23$^\circ$C, respectively. 

\begin{figure}[htb!]
	\centering
	\includegraphics[scale=0.63875]{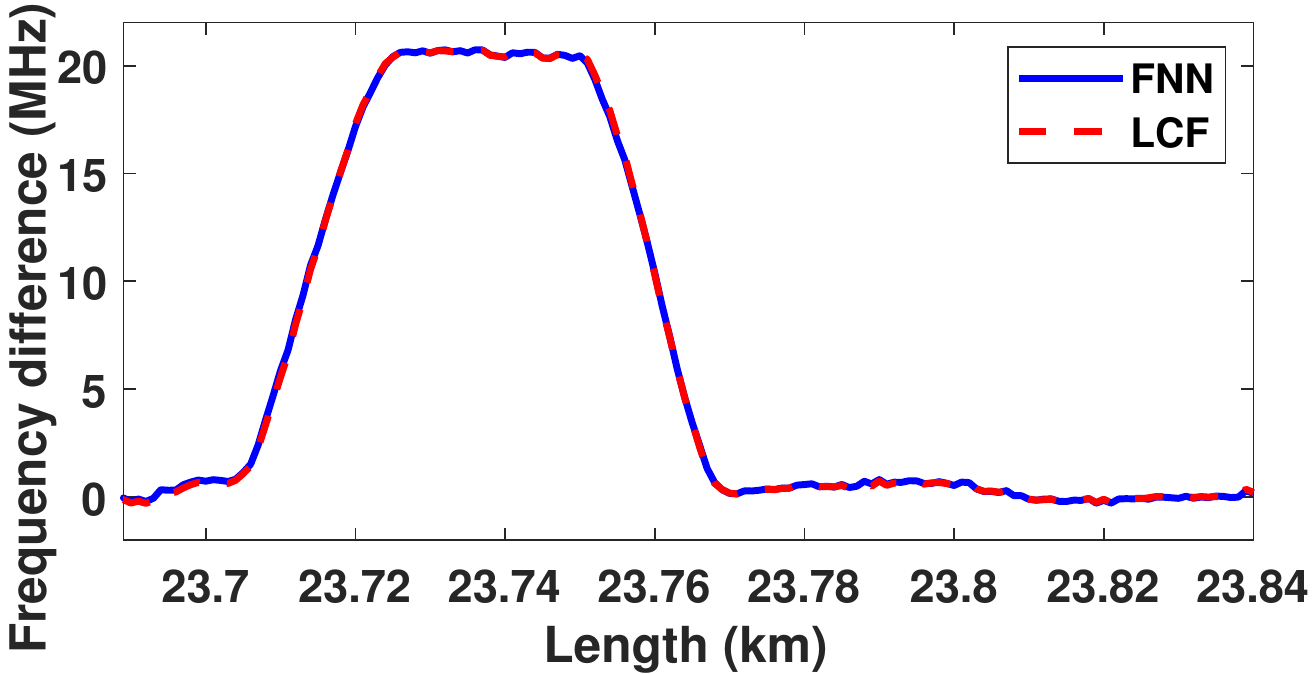}
	\caption{The frequency difference around the heating location.}
	\label{fig12}
\end{figure}

Themean deviation of the frequency difference of BFS calculated by FNN and LCF  is analyzed every 1 km, as shown in Fig. 13. 
Compared to  LCF, it indicates that FNN can accurately calculate BFS from the BGS.

\begin{figure}[htb!]
	\centering
	\includegraphics[scale=0.63875]{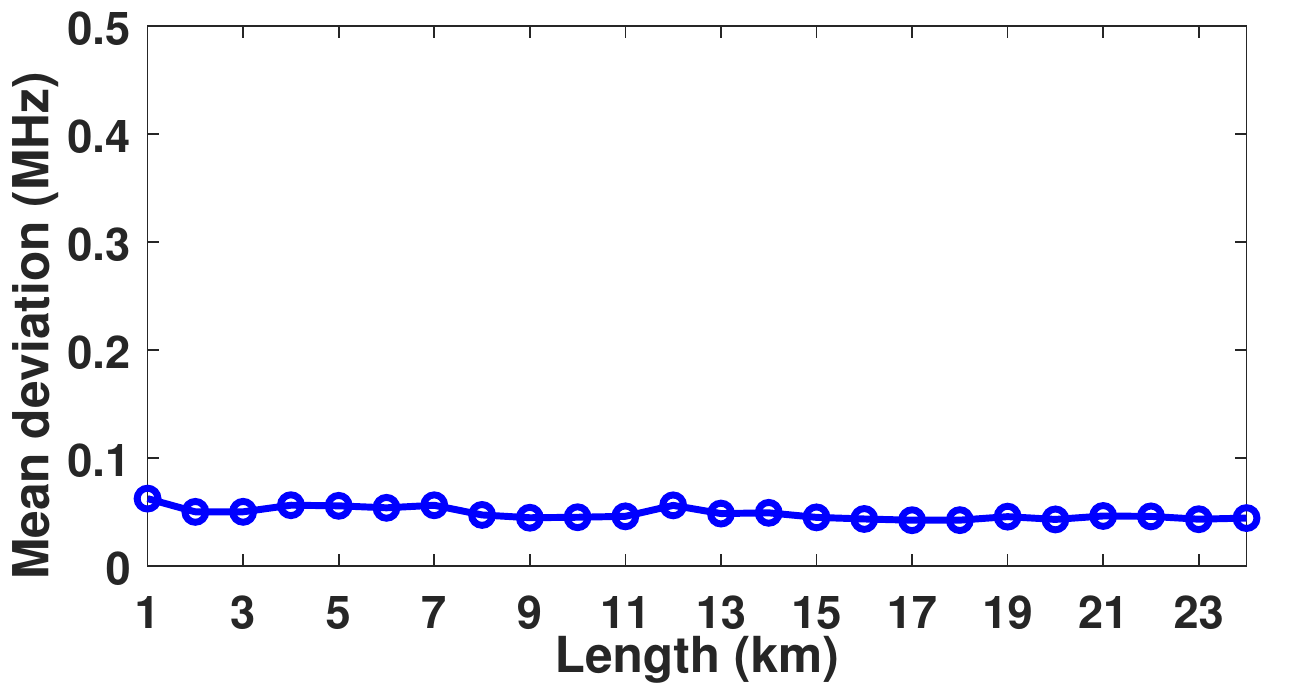}
	\caption{The mean deviation (every 1km) of the caculated BFS between FNN and LCF in the 23.95 km BOTDA.}
	\label{fig13}
\end{figure}

\subsection{Experimental setup of the 150.62 km BOTDA}
To further analyze the performance of the FNN, an advanced BOTDA of 150.62 km is built. 
The spatial resolution is 9 m, which means sixteen thousand sensing units are fused along the fiber to sense the change of temperature. 
A variety of technologies, such as hybrid distributed amplification, frequency division multiplexing (FDM), wavelength division multiplexing (WDM), time division multiplexing (TDM), are combined and used in the experiment\cite{11}. 
The experimental setup is showed in Fig. 14. 
The frequency scanning step of the BGS is 4 MHz and the frequency scanning range is 156 MHz.

\begin{figure}[htb!]
	\centering
	\includegraphics[scale=0.63875]{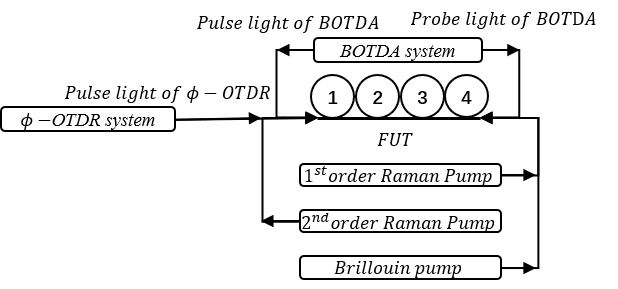}
	\caption{Experimental setup of 150.62 km BOTDA.}
	\label{fig14}
\end{figure}

The SNR of the BGS of the four-segment fiber is analyzed. 
The first three segments have the same SNR, about 27 dB, and the last segment is about 18 dB. 

Since the frequency scanning step is 4 MHz, linear interpolation is required to reshape the BGS before it is input to the FNN. 
The frequency difference of BFS calculated by FNN is compared with  LCF, as shown in Fig. 15. 
The temperature coefficient is 1.0 MHz/℃ and the applied temperature difference is 18.2 ℃. 
The temperature difference measured by LCF and FNN are both 18.7 ℃. 
The measurement uncertainty calculated by LCF and FNN are ±0.82 ℃ and ±0.83 ℃, respectively. T
he measurement uncertainty is higher than  the 23.95  km BOTDA at a frequency scanning step of 1 MHz.

\begin{figure}[htb!]
	\centering
	\includegraphics[scale=0.63875]{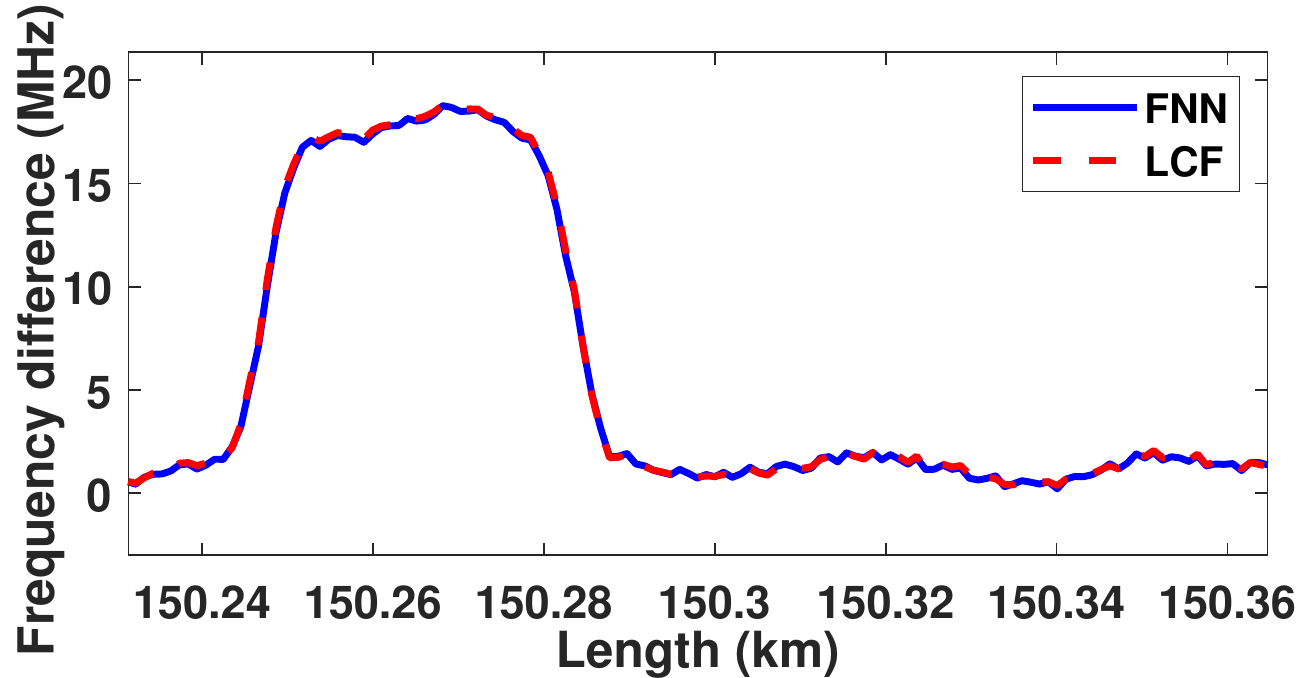}
	\caption{The frequency difference around the heating location.}
	\label{fig15}
\end{figure}

Similarly, the mean deviation of the frequency difference of BFS calculated by FNN and LCF is analyzed every 1 km, as shown in Fig. 16. 
It is acceptable that the mean deviation is less than 0.2 MHz compared with the measurement uncertainty.

\begin{figure}[htb!]
	\centering
	\includegraphics[scale=0.63875]{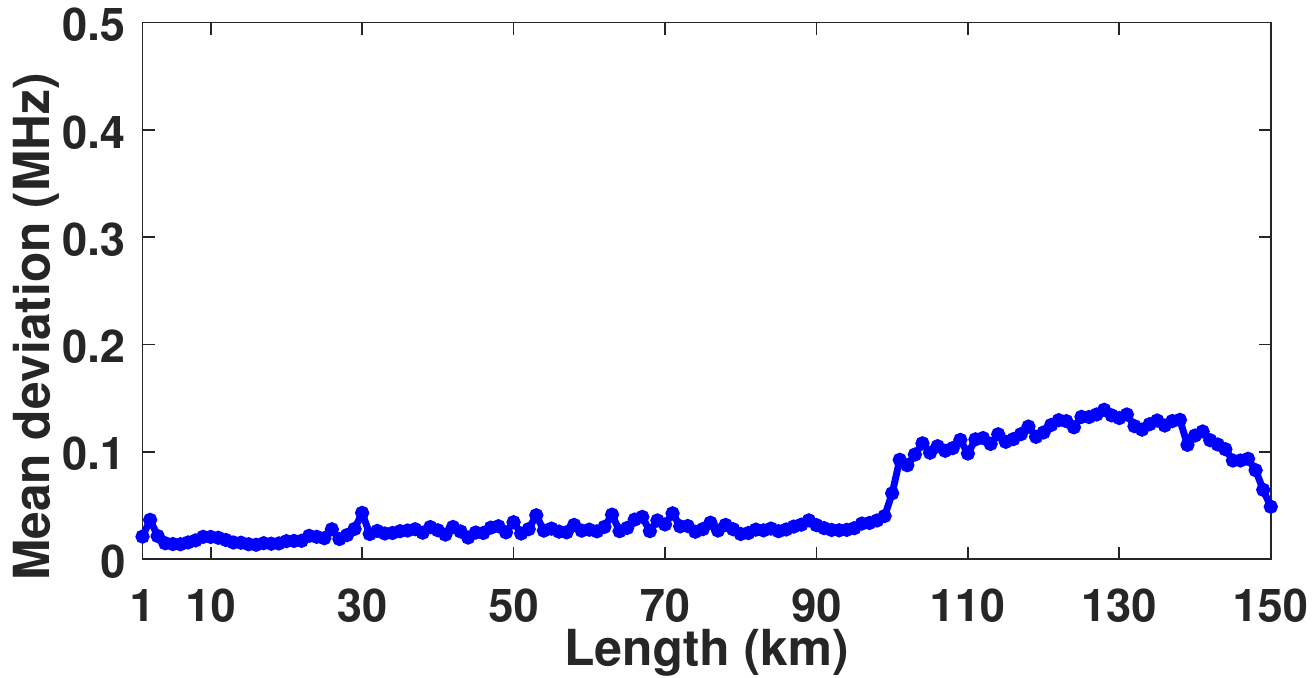}
	\caption{The mean deviation (every 1km) of the caculated BFS between FNN and LCF in the 150.62 km BOTDA.}
	\label{fig16}
\end{figure}

By constructing and analyzing BOTDA experiments, it confirms that FNN can retrieve BFS accurately under different experimental parameters.

\section{Conclusion}
Using the proposed FNN training method, BFS can be retrieved from the BGSs under various experimental parameters. 
Compared with  LCF, FNN trained with the proposed method is much more efficient, without sacrificing the measurement uncertainty. 
This work proves the value of FNN in DOFS system and promotes the development of DOFS for real-world applications.

\bibliography{ref10}

\begin{thebibliography}{10}
\providecommand{\url}[1]{#1}
\csname url@samestyle\endcsname
\providecommand{\newblock}{\relax}
\providecommand{\bibinfo}[2]{#2}
\providecommand{\BIBentrySTDinterwordspacing}{\spaceskip=0pt\relax}
\providecommand{\BIBentryALTinterwordstretchfactor}{4}
\providecommand{\BIBentryALTinterwordspacing}{\spaceskip=\fontdimen2\font plus
\BIBentryALTinterwordstretchfactor\fontdimen3\font minus
  \fontdimen4\font\relax}
\providecommand{\BIBforeignlanguage}[2]{{%
\expandafter\ifx\csname l@#1\endcsname\relax
\typeout{** WARNING: IEEEtran.bst: No hyphenation pattern has been}%
\typeout{** loaded for the language `#1'. Using the pattern for}%
\typeout{** the default language instead.}%
\else
\language=\csname l@#1\endcsname
\fi
#2}}
\providecommand{\BIBdecl}{\relax}
\BIBdecl

\bibitem{1}
A.~Zanella, N.~Bui, A.~Castellani, L.~Vangelista, and M.~Zorzi, ``{Internet of
  Things for Smart Cities},'' \emph{IEEE Internet of Things Journal}, vol.~1,
  no.~1, pp. 22--32, Feb. 2014.

\bibitem{2}
E.~Hodo, X.~Bellekens, A.~Hamilton, P.~Dubouilh, E.~Iorkyase, C.~Tachtatzis,
  and R.~Atkinson, ``{Threat analysis of IoT networks using artificial neural
  network intrusion detection system},'' in \emph{2016 International Symposium
  on Networks, Computers and Communications (ISNCC), Hammamet,Tunisia}, May.
  2016, pp. 1--6.

\bibitem{3}
H.~Cai, B.~Xu, L.~Jiang, and A.~V. Vasilakos, ``{IoT-Based Big Data Storage
  Systems in Cloud Computing: Perspectives and Challenges},'' \emph{IEEE
  Internet of Things Journal}, vol.~4, no.~1, pp. 75--87, Feb. 2017.

\bibitem{4}
T.~Yashiro, S.~Kobayashi, N.~Koshizuka, and K.~Sakamura, ``{An Internet of
  Things (IoT) architecture for embedded appliances},'' in \emph{2013 IEEE
  Region 10 Humanitarian Technology Conference, Sendai, Japan}, Aug. 2013, pp.
  314--319.

\bibitem{5}
J.~Gubbi, R.~Buyya, S.~Marusic, and M.~Palaniswami, ``{Internet of Things
  (IoT): A vision, architectural elements, and future directions},''
  \emph{Future Generation Computer Systems}, vol.~29, no.~7, pp. 1645--1660,
  Sept. 2013.

\bibitem{6}
D.~A. Krohn, T.~MacDougall, and A.~Mendez, \emph{{Fiber optic sensors:
  fundamentals and applications}}.\hskip 1em plus 0.5em minus 0.4em\relax in
  SPIE Press, 4th ed. Bellingham, WA, USA: SPIE, 2014.

\bibitem{7}
\BIBentryALTinterwordspacing
P.~Ferdinand, ``{The evolution of optical fiber sensors technologies during the
  35 last years and their applications in structure health monitoring},'' in
  \emph{EWSHM-7th European Workshop on Structural Health Monitoring}, Jul.
  2014. [Online]. Available: \url{https://hal.inria.fr/hal-01021251}
\BIBentrySTDinterwordspacing

\bibitem{8}
\BIBentryALTinterwordspacing
Z.~Wang, B.~Zhang, J.~Xiong, Y.~Fu, S.~Lin, J.~Jiang, Y.~Chen, Y.~Wu, Q.~Meng,
  and Y.~Rao, ``{Distributed acoustic sensing based on pulse-coding
  phase-sensitive OTDR},'' \emph{IEEE Internet of Things Journal}, Sept. 2018.
  [Online]. Available: \url{https://dx.doi.org/10.1109/JIOT.2018.2869474}
\BIBentrySTDinterwordspacing

\bibitem{9}
\BIBentryALTinterwordspacing
{\.I}.~{\"O}l{\c{c}}er and A.~{\"O}nc{\"u}, ``{On the use and the performance
  of adaptive filters for fiber optic distributed acoustic sensors},'' in
  \emph{2018 26th Signal Processing and Communications Applications Conference
  (SIU)}.\hskip 1em plus 0.5em minus 0.4em\relax IEEE, May 2018. [Online].
  Available: \url{https://dx.doi.org/10.1109/SIU.2018.8404578}
\BIBentrySTDinterwordspacing

\bibitem{10}
\BIBentryALTinterwordspacing
C.~Franciscangelis, F.~Fruett, W.~Margulis, L.~Kjellberg, and C.~Floridia,
  ``{Real-time multiple machines sound listening using a phase-OTDR based
  distributed microphone},'' in \emph{Microwave and Optoelectronics Conference
  (IMOC), 2017 SBMO/IEEE MTT-S International}.\hskip 1em plus 0.5em minus
  0.4em\relax IEEE, Aug. 2017. [Online]. Available:
  \url{https://dx.doi.org/10.1109/IMOC.2017.8121039}
\BIBentrySTDinterwordspacing

\bibitem{11}
Y.~Fu, Z.~Wang, R.~Zhu, N.~Xue, J.~Jiang, C.~Lu, B.~Zhang, L.~Yang, D.~Atubga,
  and Y.~Rao, ``{Ultra-Long-Distance Hybrid BOTDA/$\Phi$-OTDR},''
  \emph{Sensors}, vol.~18, no. 4, 976, Mar. 2018.

\bibitem{12}
D.~Zhou, Y.~Dong, B.~Wang, C.~Pang, D.~Ba, H.~Zhang, Z.~Lu, H.~Li, and X.~Bao,
  ``{Single-shot BOTDA based on an optical chirp chain probe wave for
  distributed ultrafast measurement},'' \emph{Light, Science \& Applications},
  vol.~7, no. 1, 32, Jul. 2018.

\bibitem{13}
S.~Lin, Z.~Wang, J.~Xiong, Y.~Fu, J.~Jiang, Y.~Wu, Y.~Chen, C.~Lu, and Y.~Rao,
  ``{Rayleigh fading suppression in one-dimension optical scatters},''
  \emph{submitted to IEEE Access}.

\bibitem{14}
M.~Nikl{\`e}s and F.~Ravet, ``{Distributed fibre sensors: Depth and
  sensitivity},'' \emph{Nature Photonics}, vol.~4, no.~7, pp. 431--432, Jul.
  2010.

\bibitem{15}
A.~Motil, A.~Bergman, and M.~Tur, ``{[INVITED] State of the art of Brillouin
  fiber-optic distributed sensing},'' \emph{Optics \& Laser Technology},
  vol.~78, pp. 81--103, Apr. 2016.

\bibitem{16}
T.~Horiguchi and M.~Tateda, ``{BOTDA-nondestructive measurement of single-mode
  optical fiber attenuation characteristics using Brillouin interaction:
  theory},'' \emph{Journal of lightwave technology}, vol.~7, no.~8, pp.
  1170--1176, Aug. 1989.

\bibitem{17}
S.~M. Haneef, Z.~Yang, L.~Th{\'e}venaz, D.~Venkitesh, and B.~Srinivasan,
  ``{Performance analysis of frequency shift estimation techniques in Brillouin
  distributed fiber sensors},'' \emph{Optics Express}, vol.~26, no.~11, pp.
  14\,661--14\,677, May. 2018.

\bibitem{18}
H.~Wu, L.~Wang, Z.~Zhao, N.~Guo, C.~Shu, and C.~Lu, ``{Brillouin optical time
  domain analyzer sensors assisted by advanced image denoising techniques},''
  \emph{Optics Express}, vol.~26, no.~5, pp. 5126--5139, Mar. 2018.

\bibitem{19}
\BIBentryALTinterwordspacing
H.~Wu, Y.~Wan, M.~Tang, Y.~Chen, C.~Zhao, R.~Liao, Y.~Chang, S.~Fu, P.~P. Shum,
  and D.~Liu, ``{Real-Time Denoising of Brillouin Optical Time Domain Analyzer
  with High Data Fidelity Using Convolutional Neural Networks},'' \emph{Journal
  of Lightwave Technology}, Oct. 2018. [Online]. Available:
  \url{https://dx.doi.org/10.1109/JLT.2018.2876909}
\BIBentrySTDinterwordspacing

\bibitem{22}
A.~K. Azad, L.~Wang, N.~Guo, H.-Y. Tam, and C.~Lu, ``{Signal processing using
  artificial neural network for BOTDA sensor system},'' \emph{Optics Express},
  vol.~24, no.~6, pp. 6769--6782, Mar. 2016.

\bibitem{23}
\BIBentryALTinterwordspacing
A.~K. Azad, L.~Wang, N.~Guo, C.~Lu, and H.~Tam, ``{Temperature profile
  extraction using artificial neural network in BOTDA sensor system},'' in
  \emph{2015 Opto-Electronics and Communications Conference (OECC)}.\hskip 1em
  plus 0.5em minus 0.4em\relax IEEE, Jun. 2015. [Online]. Available:
  \url{https://dx.doi.org/10.1109/OECC.2015.7340143}
\BIBentrySTDinterwordspacing

\bibitem{24}
\BIBentryALTinterwordspacing
B.~Wang, N.~Guo, F.~N. Khan, A.~K. Azad, L.~Wang, C.~Yu, and C.~Lu,
  ``{Extraction of temperature distribution using deep neural networks for
  BOTDA sensing system},'' in \emph{2017 Conference on Lasers and
  Electro-Optics Pacific Rim (CLEO-PR)}.\hskip 1em plus 0.5em minus 0.4em\relax
  IEEE, Jul. 2017. [Online]. Available:
  \url{https://dx.doi.org/10.1109/CLEOPR.2017.8118961}
\BIBentrySTDinterwordspacing

\bibitem{25}
A.~P. Piotrowski and J.~J. Napiorkowski, ``{A comparison of methods to avoid
  overfitting in neural networks training in the case of catchment runoff
  modelling},'' \emph{Journal of Hydrology}, vol. 476, pp. 97--111, Jan. 2013.

\bibitem{26}
N.~Murata, S.~Yoshizawa, and S.-i. Amari, ``{Network information
  criterion-determining the number of hidden units for an artificial neural
  network model},'' \emph{IEEE Transactions on Neural Networks}, vol.~5, no.~6,
  pp. 865--872, Nov. 1994.

\bibitem{27}
R.~Gencay and M.~Qi, ``{Pricing and hedging derivative securities with neural
  networks: Bayesian regularization, early stopping, and bagging},'' \emph{IEEE
  Transactions on Neural Networks}, vol.~12, no.~4, pp. 726--734, Jul. 2001.

\bibitem{28}
C.~M. Bishop, ``{Training with noise is equivalent to Tikhonov
  regularization},'' \emph{Neural Computation}, vol.~7, no.~1, pp. 108--116,
  Jan. 1995.

\end{thebibliography}


\end{document}